\documentclass[mathleft]{an}
\usepackage{graphicx}
\usepackage{times}
\usepackage{caption}
\usepackage{natbib}
\bibpunct{(}{)}{;}{a}{}{,}
\def\mnras{{MNRAS}}
\def\apj{{ApJ}}

\def\aap{{A\&A}}
\def\apjl{{ApJL}}

\def\pasp{{PASP}}

\def\mch{M$\rm^{c}$Hardy\,}
\def\me{{$\dot{m}_{E}$}}
\def\ecs{ergs cm$^{-2}$ s$^{-1}$~}

\def\msun{$M_{\odot}$}
\begin{document}

\Pagespan{1}{}
\Yearpublication{2016}%
\Yearsubmission{2015}%
\Month{10}%
\Volume{**}%
\Issue{**}%

\title{The Origin of UV-optical Variability in AGN and Test of Disc
  Models:\\
XMM-Newton and ground based observations of NGC4395}

\author{I.M. M$\rm^{c}$Hardy\inst{1}\fnmsep\thanks{Corresponding author:
  \email{imh@soton.ac.uk}\newline}
S. D. Connolly \inst{1}, B. M. Peterson\inst{2}, A. Bieryla\inst{3}, H. Chand\inst{4},
M.S. Elvis\inst{3}, D. Emmanoulopoulos\inst{1}, E. Falco\inst{3},
P. Gandhi\inst{1}, S. Kaspi\inst{5}, D. Latham\inst{3}, P. Lira\inst{6},
C. McCully\inst{7},  H. Netzer\inst{5} \and M. Uemura\inst{8}
}
\titlerunning{UV-optical variability in AGN}
\authorrunning{I.M. M$\rm^{c}$Hardy et al. }
\institute{
Department of Phyiscs and Astronomy, University of Southampton,
University Road, Southampton SO17 1BJ, UK.
\and 
LCOGT and Department of Astronomy, Ohio State University, 140 West 18th Avenue,
Columbus, OH 43210-1173
\and 
Harvard Smithsonian Center for Astrophysics, 60 Garden Street,
Cambridge, MA02138, USA.
\and
Aryabhatta Research Institute of Observational Sciences (ARIES), Manora Peak, Nainital 263 002, India
\and
Wise Observatory and School of Physics and Astronomy, Tel Aviv University,
Tel Aviv 69978, Israel
\and
Departmento de Astronomia, Universidad de Chile, Camino del
Observatorio 1515, Santiago, Chile
\and
LCOGT and Department of Physics and Astronomy, Rutgers University, 136
Frelinghuysen Road, Piscataway, NJ 0885,  USA 
\and
Hiroshima Astrophysical Science Center, Hiroshima University,
Kagamiyama 1-3-1, Hiroshima 739-8526, Japan
}

\received{October 2015}
\accepted{November 2015}
\publonline{later}

\keywords{AGN; X-rays; UV; Optical; Variability; Accretion discs}

\abstract{The origin of short timescale (weeks/months) variability of
  AGN, whether due to intrinsic disc variations or reprocessing of
  X-ray emission by a surrounding accretion disc, has been a puzzle
  for many years. However recently a number of observational
  programmes, particularly of NGC5548 with Swift, have shown that the
  UV/optical variations lag behind the X-ray variations in a manner
  strongly supportive of X-ray reprocessing. Somewhat
  surprisingly the implied size of the accretion disc is
  $\sim3 \times$ greater than expected from a standard, smooth,
  Shakura-Sunyaev thin disc model. Although the difference may be
  explained by a clumpy accretion disc, it is not clear whether the
  difference will occur in all AGN or whether it may change as, eg, a
  function of black hole mass, accretion rate or disc
  temperature. Measurements of interband lags for most AGN require
  long timescale monitoring, which is hard to arrange. However for low
  mass ($<10^{6}$\msun) AGN, the combination of XMM-Newton EPIC
  (X-rays) with the optical monitor in fast readout mode allows an
  X-ray/UV-optical lag to be measured within a single long
  observation. Here we summarise previous related observations and
  report on XMM-Newton observations of NGC4395 (mass $100 \times$
  lower, accretion rate $\sim 20 \times$ lower than for NGC5548).  We
  find that the UVW1 lags the X-rays by $\sim470$s. Simultaneous
  observations at 6 different ground based observatories also allowed
  the g-band lag ($\sim800s$) to be measured. These observations are
  in agreement with X-ray reprocessing but initial analysis suggests
  that, for NGC4395, they do not differ markedly from the predictions
  of the standard thin disc model.}

\maketitle

\section{Models for UV/Optical Variability and relationship to
  X-ray Variability}

The origin of UV/optical variability in AGN and its relationship to
X-ray variability has been a puzzle for some time and there are two
main possibilities for the origin of the UV/optical variability. The
UV/optical variability could result from reprocessing of X-ray
emission by the accretion disc or it could simply be the result of
intrinsic variability of the thermal emission from the disc. These two
models can, in principle, be distinguished simply by measuring the lag
between the X-ray and UV/optical wavebands. In the reprocessing model,
the UV/optical variations will lag behind the X-ray variations by the
light travel time between the two emission regions. For a typical AGN
this time will be a few hours. If the UV/optical variations are
produced by intrinsic disc variations there are two possible lag
timescales. If the UV/optical photons are the seed photons for the
X-ray emission, being Compton up-scattered in the central corona, then
if the X-ray variations are driven by seed photon variations, the
X-ray emission will lag behind the UV/optical variations by the light
travel time between the two emission regions, ie a few
hours. Alternatively, if the UV/optical variations are caused by
inwardly propagating accretion rate variations \citep{arevalouttley06},
these variations will eventually propagate in at the viscous timescale
and will affect the X-ray emission region. In this case the X-ray variations
will lag the UV/optical varations by $\sim$years.

\section{Previous Related Observations }

\subsection{UV/Optical inter-band lags}
The thin disc model which has been our accepted model for the
temperature structure of optically thick accretion discs for over 40
years \cite[][SS]{shakura73} predicts a disc temperature profile
$T \propto R^{-3/4}$. Therefore, in the X-ray reprocessing model, in
which incident X-ray emission boosts the existing thermal emission
from the disc, we expect that the time lag of the UV/optical
variations after the X-ray variations should be given by
$lag \propto wavelength (\lambda)^{4/3}$. 
\cite{sergeev05} and \cite{cackett07} have measured the lags of the V,
R, R1 and I bands relative to the B band for a sample of
AGN. \citeauthor{cackett07} find that, although not a perfect fit, the lags
are broadly consistent with the prediction of a reprocessing
model. However there were no accompanying X-ray measurements so it is
unknown as to how the X-ray variations might be related to the optical
variations.


\subsection{RXTE and ground based optical observations}
\begin{figure}[h]
\includegraphics[width=50mm,height=80mm,angle=270]{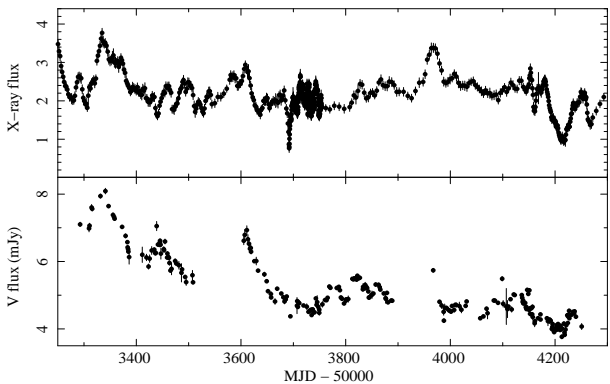}
\caption{RXTE X-ray (upper panel) and ground based V-band (lower
  panel) lightcurves of Mkn79 from \protect\cite{breedt09}.}
\label{79lcs}
\end{figure}

To investigate the link between the X-ray emission and the UV/optical
emission in AGN a number of groups \cite[e.g.][]{uttley03_5548, suganuma06,
  arevalo08_2251, arevalo09, breedt09, breedt10, breedt10_thesis,
  lira11, lira15} have monitored AGN quasi-simultaneously in X-rays
with RXTE and in optical wavebands from the ground (eg Fig.~\ref{79lcs}).
\begin{figure}[h]
\includegraphics[width=80mm,height=40mm]{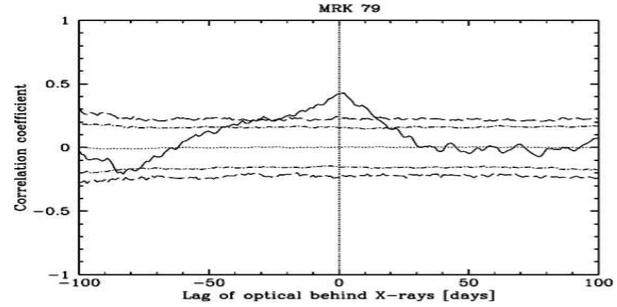}
\caption{Interpolation cross-correlation function
  \citep{white_peterson94} between the X-ray and V-band lightcurves
  shown in Fig.~\ref{79lcs}. The V-band lags the X-rays by $\sim$1d
  \protect\cite[from][]{breedt09} Here, and throughout this paper, a
  positive lag means that the longer wavelength lags behind the
  shorter wavelength.}
\label{79ccf}
\end{figure}

Cross-correlation analysis, in all cases, shows either that the
optical lags the X-rays by $\sim1$d (e.g. Fig.~\ref{79ccf}), or that
there is no measurable lag.  However the average sampling was $\sim$2d
and so the lags were rarely measured to better than 0.5d and it was
not possible to be absolutely certain that the X-rays never led the
optical.

\subsection{XMM-Newton and Swift single band lags}
To refine the measurement of the X-ray/optical lag better sampling is
needed than was available with the RXTE and ground based optical
monitoring. Such sampling is possible with Swift and with XMM-Newton.
\begin{figure}[h]
\includegraphics[width=80mm,height=40mm]{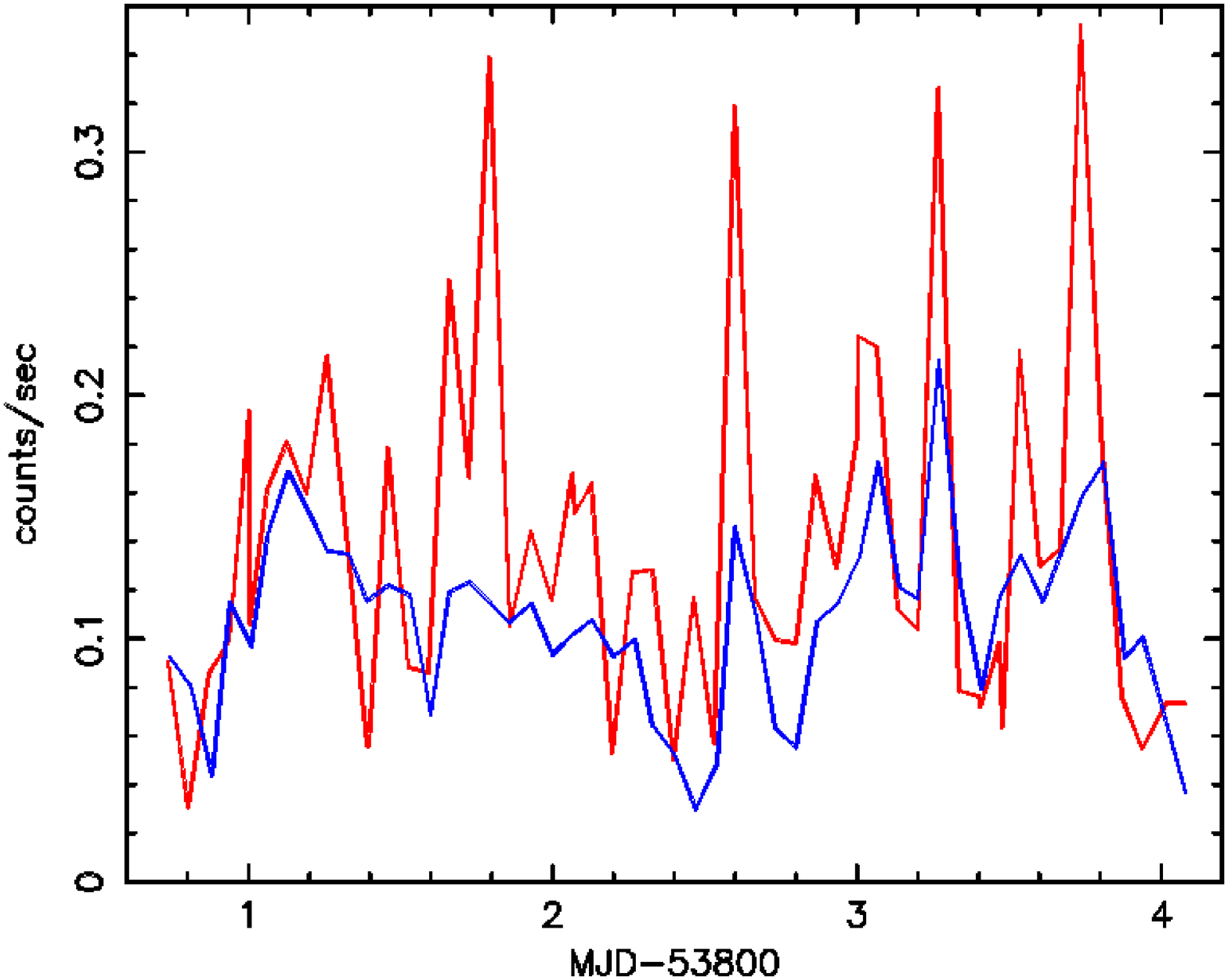}
\caption{Swift X-ray and B-band lightcurves of NGC4395
  \protect\cite[from][]{cameron12}.}
\label{4395lcs}
\end{figure}
\begin{figure}[h]
\includegraphics[width=80mm,height=40mm]{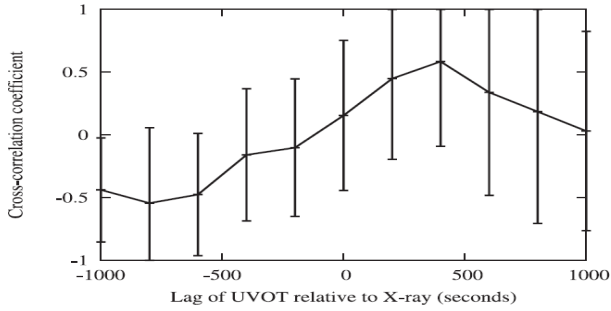}
\caption{Discrete correlation function \citep[DCF][]{edelson88}
  derived from Swift event-mode UVOT data and X-ray photon counting
  data.  The UVW2 lags behind the X-rays by $\sim$400s, although at
  low significance \protect\cite[from][]{cameron12}}.
\label{4395ccf}
\end{figure}
\begin{figure}[h]
\includegraphics[width=78mm,height=40mm]{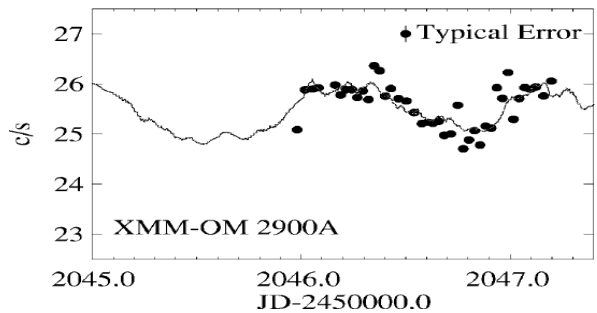}
\caption{XMM-Newton OM UVW1 imaging observations (black dots) of NGC4051 with 1ks exposure superposed on a model X-ray lightcurve from EPIC and RXTE
  observations, reprocessed by a ring of radius 0.14d
  \protect\cite[from][]{mason02}.}
\label{4051om}
\end{figure}
Swift allows observations in the 0.5-10 keV X-ray band with the X-Ray
Telescope (XRT) and, with the UV-Optical Telescope (UVOT),
simultaneous observations in one of either 3 UV (UVW2, UVM2 and UVW1)
or 3 optical (U, B and V) bands, depending on filter
selection. XMM-Newton allows X-ray observations with EPIC in a
similar band to Swift, and also allows UV/optical observations with
the Optical Monitor (OM), which is similar to the Swift UVOT.

In Fig.~\ref{4395lcs} we show Swift X-ray and B-band variations in the low
black hole mass AGN NGC4395 (mass $3.6 \times 10^{5}$\msun - see Bentz
and Katz http://www.astro.gsu.edu/AGNmass/ for all masses and
luminosities). The data here are presented with a resolution of one
satellite orbit (96 minutes).
We can see a strong correlation and cross-correlation
analysis (not shown here) reveals that the B-band lags the X-rays by
less than the orbital sampling time \citep{cameron12}.

To obtain still higher time resolution it is possible to
split the Swift orbital observations, of total duration
$\sim$1000-1500s, into smaller time bins, eg 100 or 200s.
In Fig.~\ref{4395ccf} we show the DCF derived from Swift X-ray and
UVW2-band observations with 200s time resolution (lightcurves not
shown here). Although with large errors,
this DCF suggests that the UVW2-band lags the X-rays by
$\sim$400s. This lag is approximately what we expect based on
X-ray reprocessing and formed the basis of the exciting XMM-Newton
observations which we describe in Section~\ref{4395xmm}.

With XMM-Newton the OM has, until our observations which we report here
in Section 3, usually been used in imaging mode. This mode provides
a minimum exposure time of 800s with a typical 300s readout time, thus
limiting time resolution to about 1100s. Observations in this mode
have not, in general, found significant correlations between the X-ray
and UV/optical bands \citep{smith07}. One exception is NGC4051
\citep{mason02} where, at $85\%$ confidence, the UVW1 band was seen to
lag the X-rays by $\sim$0.2d (Fig.~\ref{4051om}), broadly
consistent with reprocessing.

\subsection{Swift multi-band lags}
A number of programmes have been undertaken with Swift to measure the
lags between the X-ray band and the 6 UVOT bands. \cite{shappee14}
presented lag measurements of the Seyfert galaxy NGC2617 but
the results are slightly puzzling. The lags increase with wavelength,
but if the fit is forced to go through the X-ray point then, for
$lag \propto \lambda^{\beta}$, they find $\beta=0.37$, which is far
from the 4/3 expected from reprocessing. If the X-ray point is
ignored, a fit of $\beta=1.18$ is found between the other bands, but
if extrapolated to the X-ray wavelength, the fit is offset from the
X-ray point by a very large lag of 2.4d.

\begin{figure}[h]
\includegraphics[width=50mm,height=80mm,angle=270]{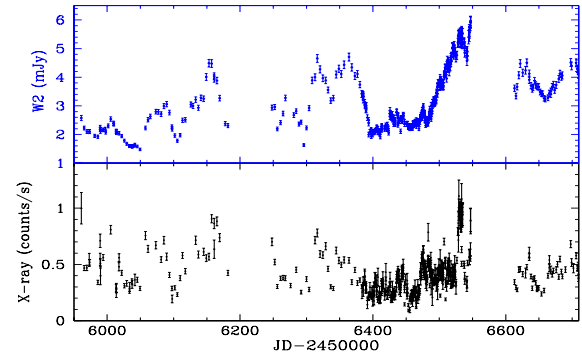}
\caption{Long timescale Swift X-ray (lower panel) and UVW2 (upper
  panel) observations of NGC5548 \protect\citep{mch14}.
}
\label{5548xw2long}
\end{figure}

\hspace*{-4mm}
\begin{minipage}{50mm}
\hspace*{-2mm}
\includegraphics[width=50mm,height=40mm,angle=0]{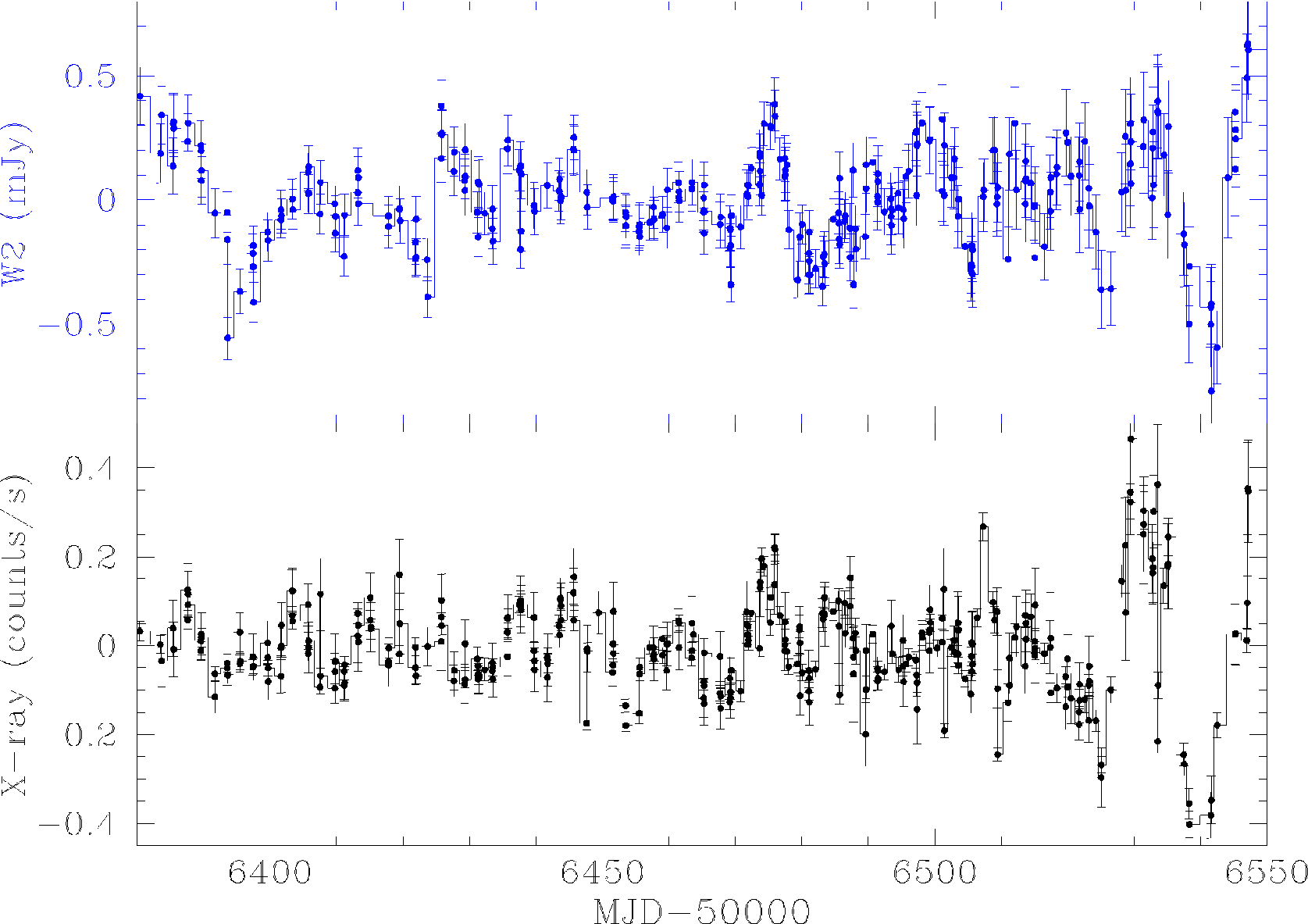}
\captionof{figure}{Swift X-ray (lower panel) and UVW2 (upper panel) 
  for a $\sim$150d period of twice-daily observations of NGC5548 
  \protect\citep{mch14} with a 20d running mean subtracted from both.}
\label{5548intensive}
\hspace*{2mm}
\end{minipage}
\hspace*{1mm}
\begin{minipage}{25mm}
\includegraphics[width=25mm,height=40mm]{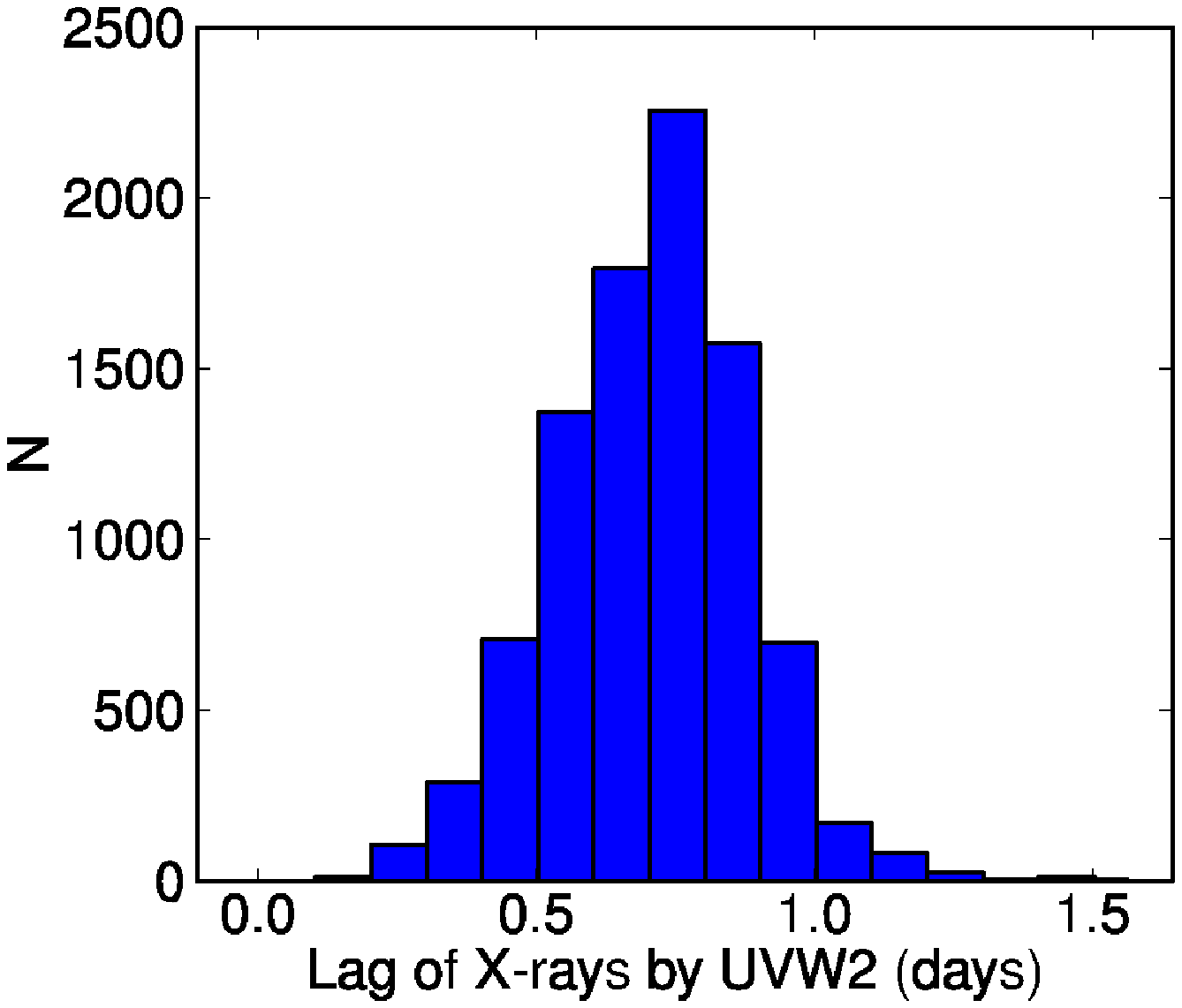}
\captionof{figure}{Lag
  of X-rays by UVW2 in NGC5548 from the
  data shown in Fig.~\ref{5548intensive}. 
}
\label{5548javelin}
\hspace*{2mm}
\end{minipage}

\cite{mch14} presented the result of almost 3 years of Swift
monitoring of NGC5548 with $\sim10 \times$ more observations than
those of \citeauthor{shappee14} In order to follow the Swift preference to
use the 'filter of the day', most of the UVOT observations were in the
UVW2 band (Fig.~\ref{5548xw2long}) which shows a strong ($>99.99\%$
confidence) correlation with the X-rays. There are, however, trends in
the UVW2 emission, lasting for a few months, which are not present in
the X-ray emission. Such trends may arise from intrinsic disc
variations caused by inwardly propagating accretion rate fluctuations.
Such long term trends, which are not present in both lightcurves, can
distort cross correlation functions and so, to measure lags on shorter
timescales, it is recommended practice to remove such trends by
subtracting a running mean \citep{welsh99}. From a 150d period
containing 300 observations \mch\, et al. therefore subtracted a 20d
running mean from both lightcurves and a strong X-ray/UVW2 correlation
is then seen (Fig.~\ref{5548intensive}). 

The lag between the two bands
was measured using a variety of techniques \citep[e.g. ZDCF,][]
{alexander13}, all showing that the UVW2 lags behind the X-rays by
about 1d. In Fig.~\ref{5548javelin} we show the lag 
($0.70^{+0.24}_{-0.27}$d) as measured using Javelin
\citep{zu11_javelin}.  These observations provided the first
unambiguous evidence that the UV variability was both strongly
correlated with the X-ray variations and lagged behind the X-ray
variations.

Javelin was designed to improve continuum-line lag measurements in
AGN. It assumes that the line lightcurve is a scaled, smoothed and
displaced version of the continuum. It models the variability as a
damped random walk (DRW), to interpolate between gaps, and directly
compares simulated line lightcurves with the observed line
lightcurve to recover the lag. \cite{pancoast14} show that
Javelin recovers simulated lags very well. It might be argued that
a DRW, which has a long timescale power spectral slope, $\alpha_{L}$,
of 0, does not describe the long timescale (months/years) X-ray
variability of AGN \cite[$\alpha_{L}= -1$, e.g.][]{mch04,mch05a} very
well. However on short timescales the DRW and X-ray power spectral
slopes are similar so there is no obvious reason why Javelin should be
less applicable here than in the measurement of continuum-line
lags.

\cite{mch14} also present observations of NGC5548 in the 5 other UVOT bands
thereby allowing the best measurement at that time of lag as a
function of wavelength (Fig.~\ref{5548lags}). In this case the fit
goes straight through the X-ray point with no offset, with
$\beta = 1.23 \pm 0.31$, in very good agreement with a reprocessing
model.

\subsection{Comparison of Lags with Models}

\begin{figure}[h]
\includegraphics[width=80mm,height=40mm]{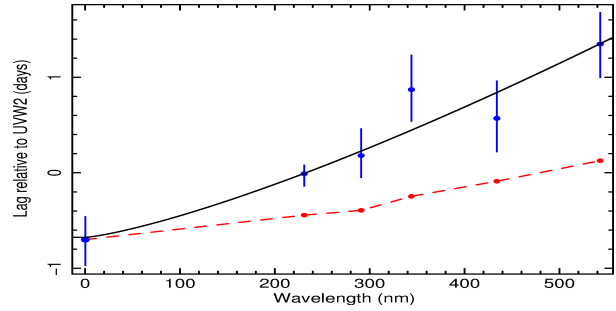}
\caption{Lags relative to the UVW2 band for NGC5548 
  \protect\cite[from][]{mch14}. The solid line is the best fit through
  all of the data, including the X-ray point. The fit of $lag \propto
  \lambda^{\beta}$ gives $\beta=1.23$, in good agreement with X-ray
  reprocessing. However the expected lags, assuming a
  standard thin disc \citep{shakura73}, are
  $\sim3\times$ shorter (dashed red line) than observed.
}
\label{5548lags}
\end{figure}

In Fig.~\ref{5548lags} we also show the expected model lags following
impulse X-ray illumination of a standard thin SS disc for the accepted
black hole mass, accretion rate and illuminating X-ray luminosity of
NGC5548.  Here the model lags are defined by the time after the
initial X-ray impulse illumination for half of the reprocessed light
to be received.  The observed lags are factors of $\sim3 \times$
larger than the model lags. Only by invoking a much hotter accretion
disc, eg by assuming a much higher accretion rate or illuminating X-ray flux
than currently accepted values, can we push the standard model close to
agreement with the observations.  Later observations of NGC5548 by
\cite{edelson15} and \cite{fausnaugh15} with extended UV and optical
monitoring confirmed this result and
\cite{troyer15} find a similar result in NGC6814. However although
initially surprising, microlensing observations \citep{morgan10} have
also indicated that accretion discs might be factors of a few larger
than predicted by SS discs.  A possible explanation is an
inhomogeneous disc temperature structure. Hotter clumps at large radii can
enhance the emission at those radii, making the disc appear larger by
factors of a few, depending on the degree of clumpiness
\citep{dexter11}.

\section{XMM-Newton and Ground Based Observations of NGC4395 }
\label{4395xmm}
As we only have good lag measurements for one AGN, it is very
important to make lag measurements on other AGN to determine
whether NGC5548 is just unusual or whether standard SS disc theory is
incomplete. Lag measurements on larger mass AGN require long
observational campaigns. However for smaller mass AGN such as NGC4395,
the expected lags can be very well measured in long observations with
XMM-Newton using EPIC for X-rays in combination with the OM in fast
readout mode for the UV/optical. The fast readout mode has not been
widely used for AGN observations but allows continuous readout with
sub-second resolution. The suggested 400s UV lag in NGC4395 cannot be
detected with standard OM imaging observations with $\sim1100$s time resolution.
On 28 and 30 December 2014 we therefore observed NGC4395 for
$\sim$53ks each time with XMM-Newton.  We observed with the OM in the
UVW1 band thus extending our coverage to shorter wavelengths than can
be observed from the ground. This band has the highest sensitivity of
the UV bands and less host galaxy contamination than the optical
bands.

\begin{figure*}[h]
\includegraphics[width=170mm,height=80mm]{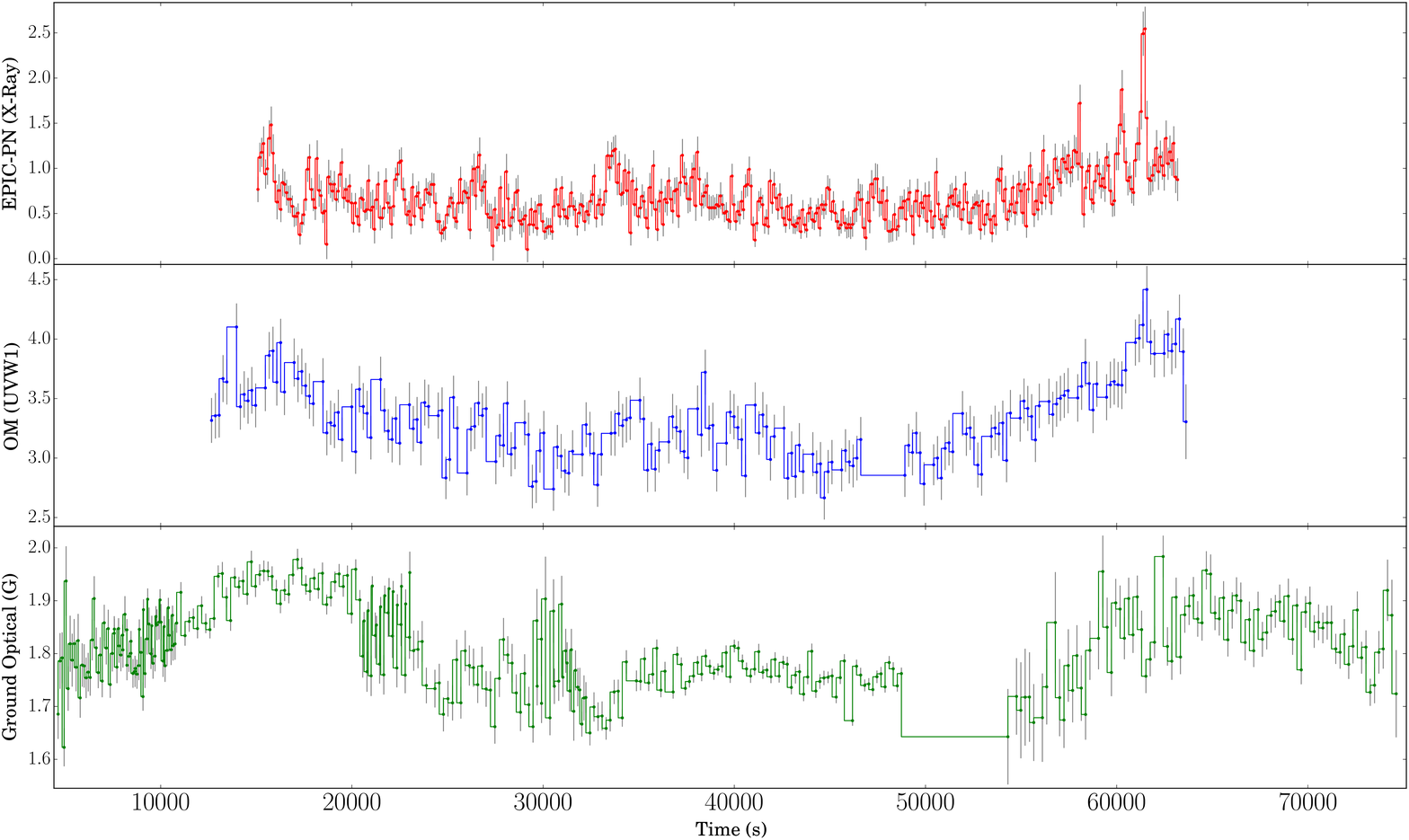}
\caption{XMM-Newton EPIC X-rays (top panel), OM UVW1 (middle panel) and ground based g-band (bottom panel) from 30 December 2014 (Connolly et al, in prep).  The X-rays are binned to 100s and the UVW1 and g-band to 200s.}
\label{xmmlcs}
\end{figure*}

During an XMM-Newton observation a source is typically
only observable from the ground for $\sim4$hr. To provide simultaneous
g-band observations we therefore made CCD imaging observations with
either 100 or 200s integrations depending on telescope size at 6
different ground based observatories (LCOGT McDonald Observatory,
Texas; Whipple Observatory, Arizona; LCOGT Haleakala Maui; Kanata
Observatory, Japan; ARIES observatory, India and the Wise Observatory,
Israel).  In Fig.~\ref{xmmlcs} we show the X-ray, UVW1 and combined
g-band lightcurves from 30 December 2014. A good correlation is seen
between all bands. In Figs.~\ref{xmmdcfw1} and ~\ref{xmmdcfg} we show
the DCFs between the X-ray band and the UVW1 and g-band lightcurves
respectively, confirming high significance correlations.

\begin{figure}[h]
\includegraphics[width=80mm,height=40mm]{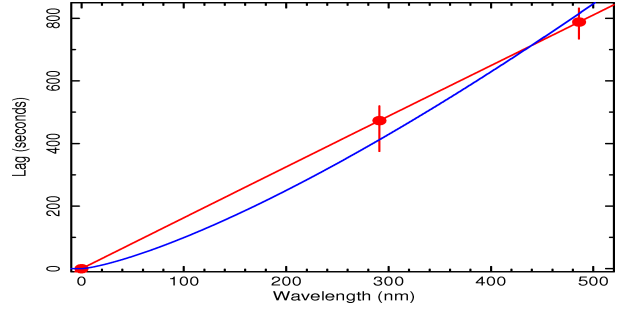}
\caption{Lags of UVW1 and g-band behind the X-rays in NGC4395 derived
  from the data shown in Fig.~\ref{xmmlcs}. For
  $lag \propto \lambda^{\beta}$, the best fit, if forced through the
  origin gives $\beta=1.0$ (red line). However $\beta=4/3$ (blue line)
  is also an
  acceptable fit.}
\label{xmmlags}
\end{figure}

\begin{figure}[h]
\includegraphics[width=78mm,height=40mm]{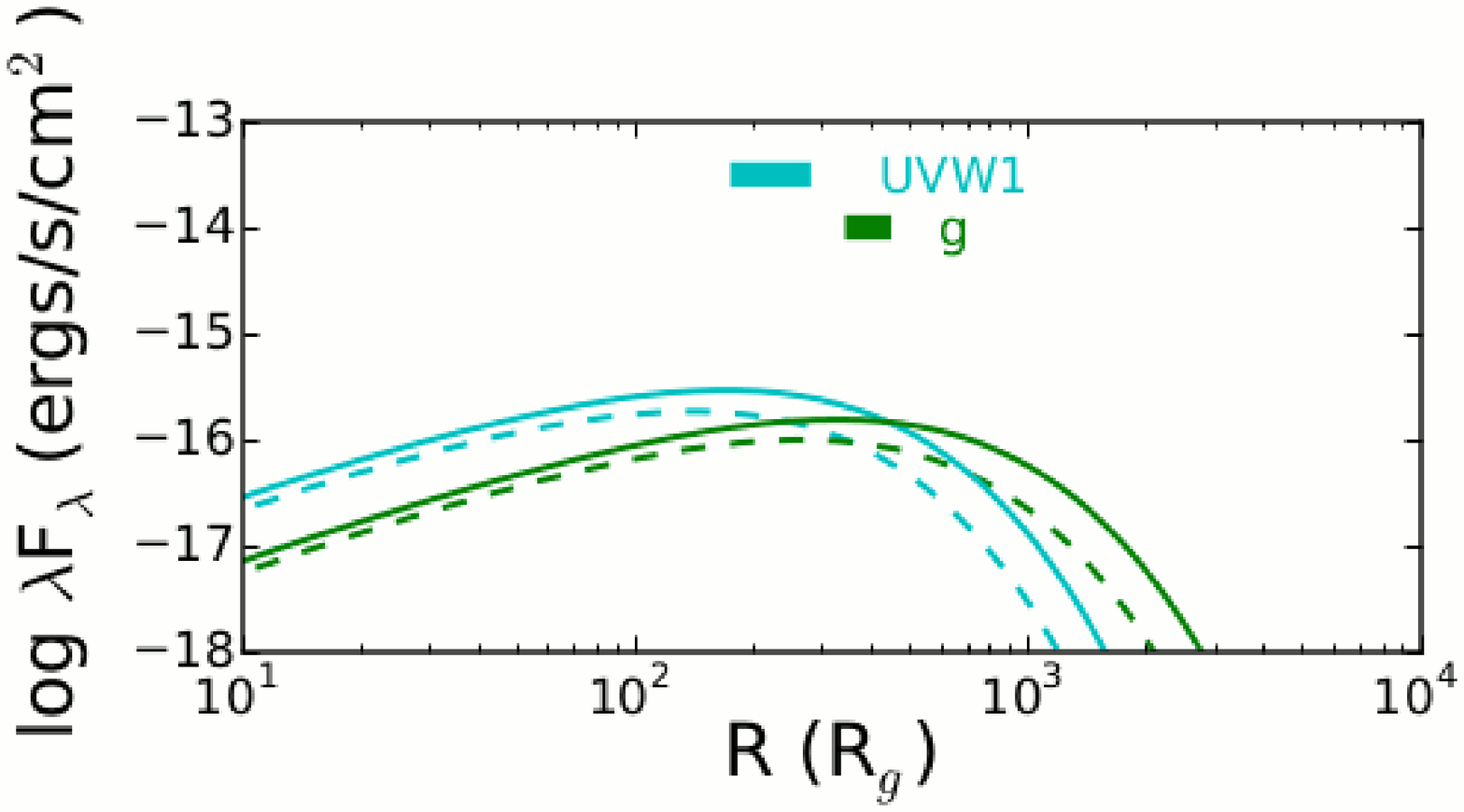}
\caption{Emissivity of the accretion disc as a function of radius in
  NGC4395 for the g-band (dark green) and UVW1 (turquiose)
  bands  \protect\cite[following][]{lira11}. We assume
$M= 3.6\times10^{5}$ \msun, \me=0.0013 and $R_{in}=3R_{g}$. We also assume 
$L_{2-10}=2.8\times10^{42}$ \ecs and extrapolate from 0.1 to 500 keV
but also assume a high albedo so that the power absorbed by the disc
is approximately equal to just $L_{2-10}$keV. The solid lines are the
total luminosity released by the disc and the dotted lines are just the
gravitational emission.}
\label{4395model}
\end{figure}

To refine the lags relative to the X-rays we calculate
lag probability distributions using Javelin for both the UVW1
(Fig.~\ref{xmmjavelinw1}) and g-bands (Fig.~\ref{xmmjaveling}). The
resultant lags are $473^{+47}_{-98}$ and $788^{+44}_{-54}$s. In
Fig.~\ref{xmmlags} we plot both the UVW1 and g-band lags as a function
of wavelength. If we force the fit through zero, a simple linear fit
(ie $\beta=1$, red line) is best although $\beta=4/3$ (blue line) is
also an acceptable fit. These observations indicate that reprocessing
of X-rays is also responsible for the UV/optical variability of
NGC4395.
\begin{minipage}{40mm}
\hspace*{-3mm}
\includegraphics[width=40mm,height=40mm]{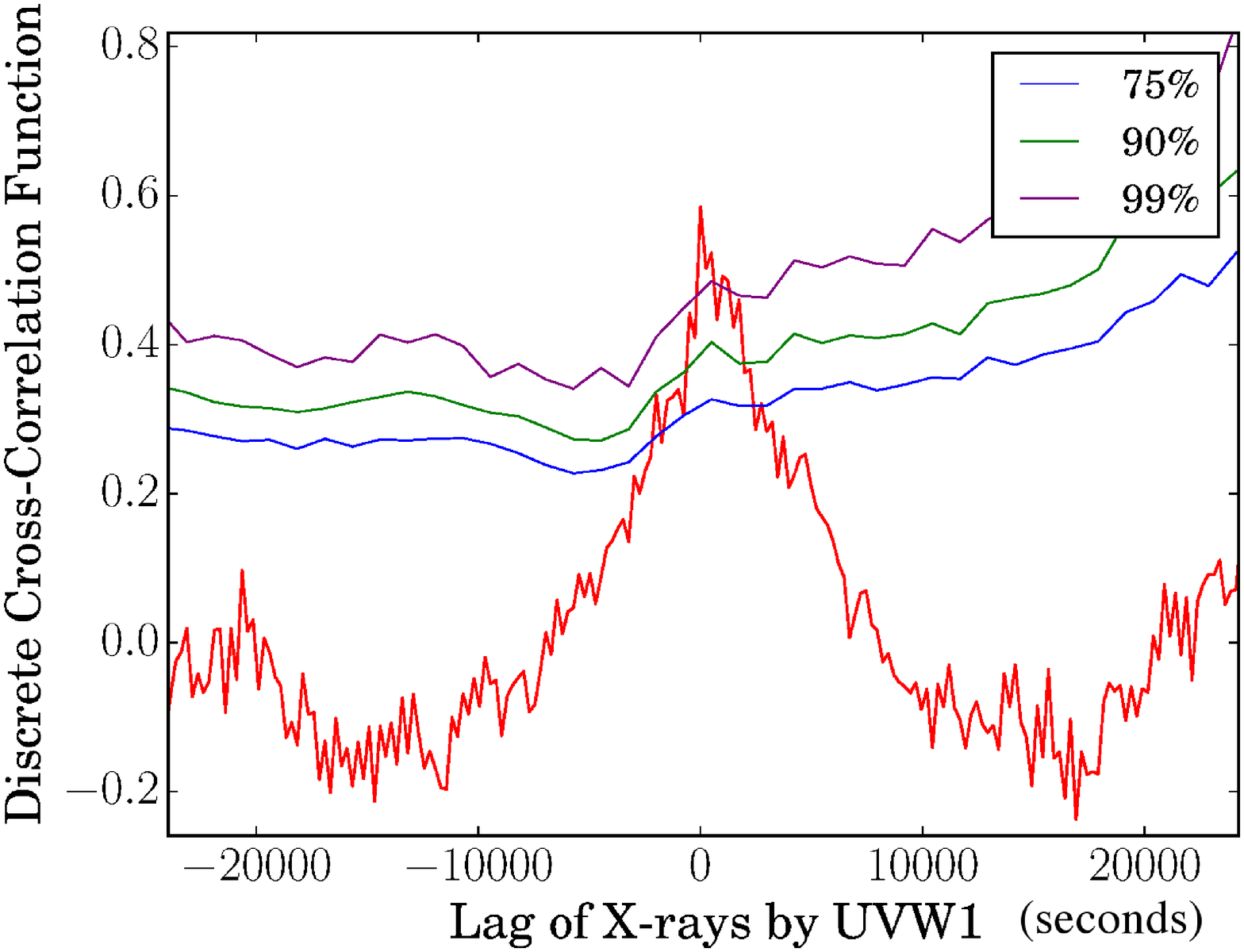}
\captionof{figure}{DCF between the X-ray and UVW1 data for NGC4395 shown in
  Fig.~\ref{xmmlcs}. Simulation based 75\%, 90\% and 99\% confidence
  levels are shown.\vspace*{10mm}}
\label{xmmdcfw1}
\end{minipage}
\hspace*{1mm}
\begin{minipage}{40mm}
\includegraphics[width=40mm,height=40mm]{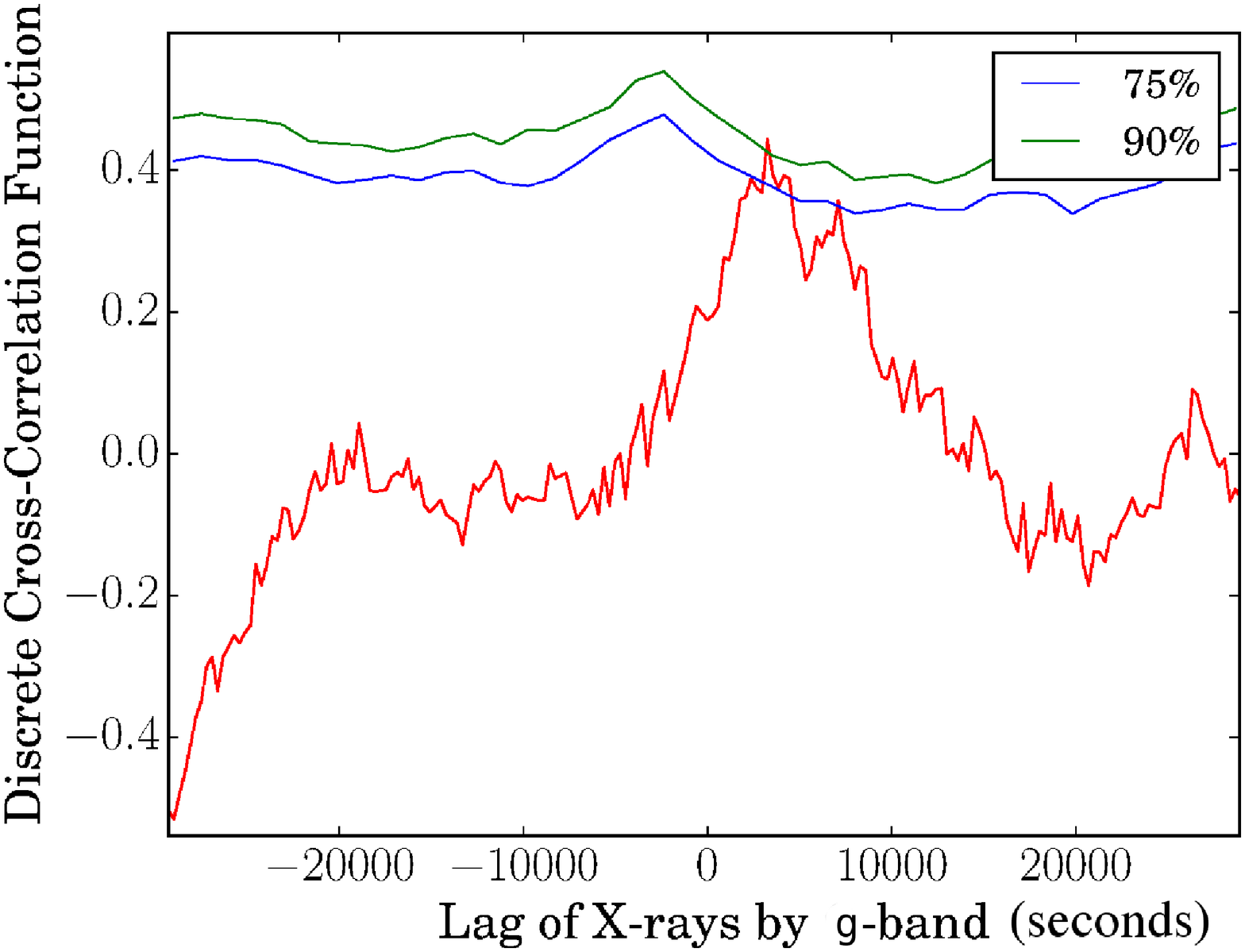}
\captionof{figure}{DCF between the X-ray and g-band data for NGC4395
  shown in Fig.~\ref{xmmlcs}.  Simulation based 75\% and 90\%
  confidence levels are shown.\vspace*{10mm} }
\label{xmmdcfg}
\end{minipage}
In Fig.~\ref{4395model} we compare the expected radial emissivity profile of
the disc with the distances derived from the lag measurements,
following \cite{lira11}. The distances derived from lag measurements
are close to the expected peak emissivity regions. Due to
computational problems we have not yet derived lags in the same way as
for NGC5548 (Fig.~\ref{5548lags}) but
 Fig.~\ref{4395model} indicates that, for NGC4395, the agreement
between standard SS thin disc theory and model may be closer than for NGC5548.

\begin{minipage}{37mm}
\hspace*{-5mm}
\includegraphics[width=37mm,height=38mm]{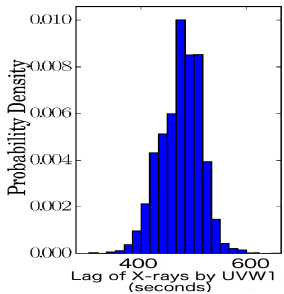}
\captionof{figure}{Lag of X-rays by UVW1
for NGC4395 from the data shown in Fig.~\ref{xmmlcs} using Javelin.}
\label{xmmjavelinw1}
\end{minipage}
\hspace*{1mm}
\begin{minipage}{38mm}
\includegraphics[width=38mm,height=40mm]{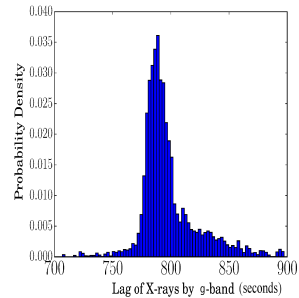}
\captionof{figure}{Lag of X-rays by g-band in NGC4395 from the data shown in Fig.~\ref{xmmlcs} using Javelin.}
\label{xmmjaveling}
\end{minipage}

\section{Conclusions }
To test whether the X-ray to UV/optical lags in an AGN which is quite
different to NGC5548 are also larger than expected from the standard
SS model, we made XMM-Newton EPIC and OM observations of NGC4395. This
AGN has a mass $\sim 100 \times$ less and accretion rate (in Eddington
units) $\sim 20 \times$ less than that of NGC5548. To obtain high
enough time resolution in the OM to sample the previously suggested
400s UV lag we used the fast readout mode, not previously used for AGN
observations. To obtain the UV lightcurve with the highest possible
signal to noise we used the UVW1 filter. To obtain a high quality,
continuous, optical lightcurve we chose the g-band and made
observations from six different observatories around the globe.

All observations were very successful and we measured lags, relative
to the X-rays of $473^{+47}_{-98}$ and $788^{+44}_{-54}$s for the UVW1
and g-bands respectively. These lags are in good agreement with the
hypothesis that the UV/optical variability is driven by reprocessing
of X-ray emission. However, unlike in NGC5548, preliminary modelling
indicates that the measured lags are not too far different from those
expected from standard SS disc theory. We remain cautious in our
interpretation at present but we do note that the disc in NGC4395 is
$\sim50\%$ hotter than in NGC5548. Increased disc
temperature may lead to a more stable disc \cite[e.g.][]{churazov01},
less sensitive to radiation-induced perturbations  \citep{pringle97}.

These observations demonstrate the  potential of XMM-Newton for
X-ray/UV lag measurements in low mass AGN.

\vspace*{4mm}
\noindent
{\bf Acknowledgements}
\noindent
We thank the XMM-Newton operations team for considerable advice and
assistance in setting up the observations of NGC4395. We also thank
the staff and observers at the McDonald, FLWO, Haleakala, Kanata,
Aries and Wise observatories. IMcH thanks STFC for support under grant
ST/M001326/1.


\begin{thebibliography}{33}
\expandafter\ifx\csname natexlab\endcsname\relax\def\natexlab#1{#1}\fi

\bibitem[{{Alexander}(2013)}]{alexander13}
{Alexander} T., 2013, ArXiv e-prints

\bibitem[{{Ar{\'e}valo} \& {Uttley}(2006)}]{arevalouttley06}
{Ar{\'e}valo} P., {Uttley} P., 2006, \mnras, 367, 801

\bibitem[{{Ar{\'e}valo} {et~al}\mbox{.}(2008){Ar{\'e}valo}, {Uttley}, {Kaspi},
  {Breedt}, {Lira}, \& {M$\rm^{c}$Hardy}}]{arevalo08_2251}
{Ar{\'e}valo} P., {Uttley} P., {Kaspi} S., {Breedt} E., {Lira} P.,
  {M$\rm^{c}$Hardy} I.~M., 2008, \mnras, 389, 1479

\bibitem[{{Ar{\'e}valo} {et~al}\mbox{.}(2009){Ar{\'e}valo}, {Uttley}, {Lira},
  {Breedt}, {M$\rm^{c}$Hardy}, \& {Churazov}}]{arevalo09}
{Ar{\'e}valo} P., {Uttley} P., {Lira} P., {Breedt} E., {M$\rm^{c}$Hardy} I.~M.,
  {Churazov} E., 2009, \mnras, 397, 2004

\bibitem[{{Breedt}(2010)}]{breedt10_thesis}
{Breedt} E., 2010, PhD thesis, Physics and Astronomy, University of Southampton

\bibitem[{{Breedt} {et~al}\mbox{.}(2009){Breedt}, {Ar{\'e}valo},
  {M$\rm^{c}$Hardy}, {Uttley}, {Sergeev}, {Minezaki}, {Yoshii}, {Gaskell},
  {Cackett}, {Horne}, \& {Koshida}}]{breedt09}
{Breedt} E. {et~al.}, 2009, \mnras, 394, 427

\bibitem[{{Breedt} {et~al}\mbox{.}(2010){Breedt}, {M$\rm^{c}$Hardy},
  {Ar{\'e}valo}, {Uttley}, {Sergeev}, {Minezaki}, {Yoshii}, {Sakata}, {Lira},
  \& {Chesnok}}]{breedt10}
{Breedt} E. {et~al.}, 2010, \mnras, 403, 605

\bibitem[{{Cackett} {et~al}\mbox{.}(2007){Cackett}, {Horne}, \&
  {Winkler}}]{cackett07}
{Cackett} E.~M., {Horne} K., {Winkler} H., 2007, \mnras, 380, 669

\bibitem[{{Cameron} {et~al}\mbox{.}(2012){Cameron}, {M$\rm^{c}$Hardy},
  {Dwelly}, {Breedt}, {Uttley}, {Lira}, \& {Arevalo}}]{cameron12}
{Cameron} D.~T., {M$\rm^{c}$Hardy} I., {Dwelly} T., {Breedt} E., {Uttley} P.,
  {Lira} P., {Arevalo} P., 2012, \mnras, 422, 902

\bibitem[{{Churazov} {et~al}\mbox{.}(2001){Churazov}, {Gilfanov}, \&
  {Revnivtsev}}]{churazov01}
{Churazov} E., {Gilfanov} M., {Revnivtsev} M., 2001, \mnras, 321, 759

\bibitem[{{Dexter} \& {Agol}(2011)}]{dexter11}
{Dexter} J., {Agol} E., 2011, \apjl, 727, L24

\bibitem[{{Edelson} {et~al}\mbox{.}(2015){Edelson}, {Gelbord}, {Horne},
  {M$\rm^{c}$Hardy}, {Peterson}, {Ar{\'e}valo}, {Breeveld}, {De Rosa}, {Evans},
  {Goad}, {Kriss}, {Brandt}, {Gehrels}, {Grupe}, {Kennea}, {Kochanek},
  {Nousek}, {Papadakis}, {Siegel}, {Starkey}, {Uttley}, {Vaughan}, {Young},
  {Barth}, {Bentz}, {Brewer}, {Crenshaw}, {Dalla Bont{\`a}}, {De
  Lorenzo-C{\'a}ceres}, {Denney}, {Dietrich}, {Ely}, {Fausnaugh}, {Grier},
  {Hall}, {Kaastra}, {Kelly}, {Korista}, {Lira}, {Mathur}, {Netzer},
  {Pancoast}, {Pei}, {Pogge}, {Schimoia}, {Treu}, {Vestergaard}, {Villforth},
  {Yan}, \& {Zu}}]{edelson15}
{Edelson} R. {et~al.}, 2015, \apj, 806, 129

\bibitem[{{Edelson} \& {Krolik}(1988)}]{edelson88}
{Edelson} R.~A., {Krolik} J.~H., 1988, \apj, 333, 646

\bibitem[{{Fausnaugh} {et~al}\mbox{.}(2015){Fausnaugh}, {Denney}, {Barth},
  {Bentz}, {Bottorff}, {Carini}, {Croxall}, {De Rosa}, {Goad}, {Horne},
  {Joner}, {Kaspi}, {Kim}, {Klimanov}, {Kochanek}, {Leonard}, {Netzer},
  {Peterson}, {Schnulle}, {Sergeev}, {Vestergaard}, {Zheng}, {Anderson},
  {Arevalo}, {Bazhaw}, {Borman}, {Boroson}, {Brandt}, {Breeveld}, {Brewer},
  {Cackett}, {Crenshaw}, {Dalla Bonta}, {De Lorenzo-Caceres}, {Dietrich},
  {Edelson}, {Efimova}, {Ely}, {Evans}, {Filippenko}, {Flatland}, {Gehrels},
  {Geier}, {Gelbord}, {Gonzalez}, {Gorjian}, {Grier}, {Grupe}, {Hall}, {Hicks},
  {Horenstein}, {Hutchison}, {Im}, {Jensen}, {Jones}, {Kaastra}, {Kelly},
  {Kennea}, {Kim}, {Korista}, {Kriss}, {Larionov}, {Lee}, {Lira}, {MacInnis},
  {Manne-Nicholas}, {Mathur}, {M$\rm^{c}$Hardy}, {Montouri}, {Musso},
  {Nazarov}, {Norris}, {Nousek}, {Okhmat}, {Pancoast}, {Papadakis}, {Parks},
  {Pei}, {Pogge}, {Pott}, {Rafter}, {Rix}, {Saylor}, {Schimoia}, {Siegel},
  {Spencer}, {Starkey}, {Sung}, {Teems}, {Treu}, {Turner}, {Uttley},
  {Villforth}, {Weiss}, {Woo}, {Yan}, {Young}, \& {Zu}}]{fausnaugh15}
{Fausnaugh} M.~M. {et~al.}, 2015, ArXiv e-prints

\bibitem[{{Lira} {et~al}\mbox{.}(2011){Lira}, {Ar{\'e}valo}, {Uttley},
  {M$\rm^{c}$Hardy}, \& {Breedt}}]{lira11}
{Lira} P., {Ar{\'e}valo} P., {Uttley} P., {M$\rm^{c}$Hardy} I., {Breedt} E.,
  2011, \mnras, 415, 1290

\bibitem[{{Lira} {et~al}\mbox{.}(2015){Lira}, {Arevalo}, {Uttley},
  {M$\rm^{c}$Hardy}, \& {Videla}}]{lira15}
{Lira} P., {Arevalo} P., {Uttley} P., {M$\rm^{c}$Hardy} I.~M.~M., {Videla} L.,
  2015, ArXiv e-prints

\bibitem[{{Mason} {et~al}\mbox{.}(2002){Mason}, {M$\rm^{c}$Hardy}, {Page},
  {Uttley}, {C{\'o}rdova}, {Maraschi}, {Priedhorsky}, {Puchnarewicz}, \&
  {Sasseen}}]{mason02}
{Mason} K.~O. {et~al.}, 2002, \apjl, 580, L117

\bibitem[{{Morgan} {et~al}\mbox{.}(2010){Morgan}, {Kochanek}, {Morgan}, \&
  {Falco}}]{morgan10}
{Morgan} C.~W., {Kochanek} C.~S., {Morgan} N.~D., {Falco} E.~E., 2010, \apj,
  712, 1129

\bibitem[{{M$\rm^{c}$Hardy} {et~al}\mbox{.}(2014){M$\rm^{c}$Hardy}, {Cameron},
  {Dwelly}, {Connolly}, {Lira}, {Emmanoulopoulos}, {Gelbord}, {Breedt},
  {Arevalo}, \& {Uttley}}]{mch14}
{M$\rm^{c}$Hardy} I.~M. {et~al.}, 2014, \mnras, 444, 1469

\bibitem[{{M$\rm^{c}$Hardy} {et~al}\mbox{.}(2005){M$\rm^{c}$Hardy}, {Gunn},
  {Uttley}, \& {Goad}}]{mch05a}
{M$\rm^{c}$Hardy} I.~M., {Gunn} K.~F., {Uttley} P., {Goad} M.~R., 2005, \mnras,
  359, 1469

\bibitem[{{M$\rm^{c}$Hardy} {et~al}\mbox{.}(2004){M$\rm^{c}$Hardy},
  {Papadakis}, {Uttley}, {Page}, \& {Mason}}]{mch04}
{M$\rm^{c}$Hardy} I.~M., {Papadakis} I.~E., {Uttley} P., {Page} M.~J., {Mason}
  K.~O., 2004, \mnras, 348, 783

\bibitem[{{Pancoast} {et~al}\mbox{.}(2014){Pancoast}, {Brewer}, \&
  {Treu}}]{pancoast14}
{Pancoast} A., {Brewer} B.~J., {Treu} T., 2014, ArXiv e-prints

\bibitem[{{Pringle}(1997)}]{pringle97}
{Pringle} J.~E., 1997, \mnras, 292, 136

\bibitem[{{Sergeev} {et~al}\mbox{.}(2005){Sergeev}, {Doroshenko},
  {Golubinskiy}, {Merkulova}, \& {Sergeeva}}]{sergeev05}
{Sergeev} S.~G., {Doroshenko} V.~T., {Golubinskiy} Y.~V., {Merkulova} N.~I.,
  {Sergeeva} E.~A., 2005, \apj, 622, 129

\bibitem[{{Shakura} \& {Sunyaev}(1973)}]{shakura73}
{Shakura} N.~I., {Sunyaev} R.~A., 1973, \aap, 24, 337

\bibitem[{{Shappee} {et~al}\mbox{.}(2014){Shappee}, {Prieto}, {Grupe},
  {Kochanek}, {Stanek}, {De Rosa}, {Mathur}, {Zu}, {Peterson}, {Pogge},
  {Komossa}, {Im}, {Jencson}, {Holoien}, {Basu}, {Beacom}, {Szczygie{\l}},
  {Brimacombe}, {Adams}, {Campillay}, {Choi}, {Contreras}, {Dietrich},
  {Dubberley}, {Elphick}, {Foale}, {Giustini}, {Gonzalez}, {Hawkins}, {Howell},
  {Hsiao}, {Koss}, {Leighly}, {Morrell}, {Mudd}, {Mullins}, {Nugent},
  {Parrent}, {Phillips}, {Pojmanski}, {Rosing}, {Ross}, {Sand}, {Terndrup},
  {Valenti}, {Walker}, \& {Yoon}}]{shappee14}
{Shappee} B.~J. {et~al.}, 2014, \apj, 788, 48

\bibitem[{{Smith} \& {Vaughan}(2007)}]{smith07}
{Smith} R., {Vaughan} S., 2007, \mnras, 375, 1479

\bibitem[{{Suganuma} {et~al}\mbox{.}(2006){Suganuma}, {Yoshii}, {Kobayashi},
  {Minezaki}, {Enya}, {Tomita}, {Aoki}, {Koshida}, \& {Peterson}}]{suganuma06}
{Suganuma} M. {et~al.}, 2006, \apj, 639, 46

\bibitem[{{Troyer} {et~al}\mbox{.}(2015){Troyer}, {Starkey}, {Cackett},
  {Bentz}, {Goad}, {Horne}, \& {Seals}}]{troyer15}
{Troyer} J., {Starkey} D., {Cackett} E., {Bentz} M., {Goad} M., {Horne} K.,
  {Seals} J., 2015, ArXiv e-prints

\bibitem[{{Uttley} {et~al}\mbox{.}(2003){Uttley}, {Edelson}, {M$\rm^{c}$Hardy},
  {Peterson}, \& {Markowitz}}]{uttley03_5548}
{Uttley} P., {Edelson} R., {M$\rm^{c}$Hardy} I.~M., {Peterson} B.~M.,
  {Markowitz} A., 2003, \apjl, 584, L53

\bibitem[{{Welsh}(1999)}]{welsh99}
{Welsh} W.~F., 1999, \pasp, 111, 1347

\bibitem[{{White} \& {Peterson}(1994)}]{white_peterson94}
{White} R.~J., {Peterson} B.~M., 1994, \pasp, 106, 879

\bibitem[{{Zu} {et~al}\mbox{.}(2011){Zu}, {Kochanek}, \&
  {Peterson}}]{zu11_javelin}
{Zu} Y., {Kochanek} C.~S., {Peterson} B.~M., 2011, \apj, 735, 80

\end{thebibliography}

\end{document}